\documentclass[aps,showpacs,preprintnumbers]{revtex4}
\usepackage{bm}

\begin{document}

\title{QED$_{4}$ Ward Identity for fermionic field in the light-front}
\author{J.H.O.Sales}
\affiliation{$^{1}$Funda\c{c}\~{a}o de Ensino e Pesquisa de Itajub\'{a} , Av. Dr. Antonio
Braga Filho, CEP 37501-002, Itajub\'{a}, MG, Brazil}
\author{A.T. Suzuki}
\affiliation{Instituto de F\'{\i}sica Te\'{o}rica-UNESP, Rua Pamplona 145, CEP\ 01405-900
S\~{a}o Paulo, SP, Brazil}
\author{J.D. Bolzan}
\affiliation{Instituto de F\'{\i}sica Te\'{o}rica-UNESP, Rua Pamplona 145, CEP\ 01405-900
S\~{a}o Paulo, SP, Brazil}
\date{\today}

\begin{abstract}

In a covariant gauge we implicitly assume that the Green's function propagates information from one point of the space-time to another, so that the Green's function is responsible for the dynamics of the relativistic particle. In the light front form, which in principle is a change of coordinates, one would expect that this feature would be preserved. In this manner, the fermionic field propagator can be split into a propagating piece and a non-propagating (``contact'') term. Since the latter (``contact'') one does not propagate information, and therefore, assumedly with no harm to the field dynamics we wanted to know what would be the impact of dropping it off. To do that, we investigated its role in the Ward identity in the light front.

\end{abstract}

\maketitle






11.10.Gh, 03.65.-w, 11.10.-z

\section{Introduction}

One of the most important concepts in quantum field theories is the question
of renormalizability. In QED (Quantum Electrodynamics) specifically, the
electric charge renormalization is guaranteed solely by the renormalization
of the photon propagator. This result is a consequence of the so-called Ward
identity, demonstrated by J.C.Ward in 1950 \cite{Ward1,Ward2,Ward3}. The
importance of this result can be seen and emphasized in the fact that without
the validity of such an identity, there would be no guarantee that the
renormalized charge of different fermions (electrons, muons, etc.) would be
the same. In other words, without such identity, charges of different
particles must have different renormalization constants, a feature not so
gratifying nor elegant. Moreover, without the Ward identity,
renormalizability would have to be laboriously checked order by order in
perturbation theory.

What the Ward identity does is to relate the vertex function of the theory
with the derivative of the self-energy function of the electron, and this
important correlation is expressed in terms of equality between the
renormalization constants, namely, $Z_{1}=Z_{2}$, where $Z_{1}$and $Z_{2}$
are the renormalization constants related to the vertex function and the
fermionic propagator respectively. Since the renormalized electric charge is
given in terms of the bare electric charge via the product $%
e_{R}=Z_{3}^{1/2}Z_{2}Z_{1}^{-1}e_{0}$, it follows immediately that $%
e_{R}=Z_{3}^{1/2}e_{0}$, i.e., electric charge renormalization depends
solely on the renormalization of the photon propagator.

We know that light-front dynamics is plagued with singularities of all sorts
and because of this the connection between the covariant quantities and
light-front quantities cannot be so easily established. If we want to
describe our theory in terms of the light-front coordinates or variables, we
must take care of the boundary conditions that fields must obey. Thus, a
simple projection from the covariant quantities to light-front quantities
via coordinate transformations is bound to be troublesome. This can be
easily seen in our checking of the QED Ward identity in the light-front,
where the fermionic propagator does bear an additional term proportional to $%
\gamma ^{+}(p^{+})^{-1}$ oftentimes called \textquotedblleft contact
term\textquotedblright\ in the literature, which, of course, is
conspicuously absent in the covariant propagator. This term, as we will see,
is crucial to the Ward identity in the light-front. The covariant
propagating term solely projected onto the light-front coordinates therefore
violates Ward identity, and therefore breaks gauge invariance. Such
result is obviously wrong and unwarranted.

The outline of our paper is as follows: We begin by considering the standard
derivation for the covariant case Ward identity and show explicitly that the
fermionic propagator there cannot be analytically regularized, otherwise
Ward identity cannot be achieved. Then we explicitly construct our fermionic
propagator in terms of the light-front coordinates, with the proper contact
term in it and in the following section we deal with the checking of the
Ward identity proper. Finally, the next two sections are devoted to the
concluding remarks and Appendix; in the latter we define our light-cone
coordinates convention and notation and include explicit calculations
showing that without the contact term in the fermionic propagator, Ward
identity is not satisfied, and thus gauge invariance is violated.

\section{The Ward Identity}

There are several ways to write down the Ward identity for
fermions, and one of them is inferred from manipulations of their
propagator, namely, $S(p)$. Multiplying by its inverse, we get the identity

\[
S(p)S^{-1}(p)=I\text{,}
\]

Deriving both sides with respect to $p^{\mu }$ we get

\[
\frac{\partial S(p)}{\partial p^{\mu }}S^{-1}(p)+S(p)\frac{\partial S^{-1}(p)%
}{\partial p^{\mu }}=0
\]

which leads to

\[
\frac{\partial S(p)}{\partial p^{\mu }}S^{-1}(p)=-S(p)\frac{\partial
S^{-1}(p)}{\partial p^{\mu }}
\]

Finally, multiplyng both sides from the left by the propagator itself

\begin{equation}
\frac{\partial S(p)}{\partial p^{\mu }}=-S(p)\frac{\partial S^{-1}(p)}{%
\partial p^{\mu }}S(p)  \label{passo}
\end{equation}

Now, using $\displaystyle {S(p)=\frac{i}{p\!\!\!/-m}}$ it follows that its inverse is 
$\displaystyle {S^{-1}(p)=-i(p\!\!\!/-m)}$. Deriving this last expression with respect to $p^{\mu }$ we get 

\[
\frac{\partial S^{-1}(p)}{\partial p^{\mu }}=-i\gamma _{\mu }
\]
which inserted into (\ref{passo}) leads to the differential form of the Ward
Identity, namely,

\begin{equation}
\frac{\partial S(p)}{\partial p^{\mu }}=iS(p)\gamma _{\mu }S(p)\text{.}
\label{ward}
\end{equation}

Here it is important to stress that if the propagator were raised to a power as in the analytic regularization scheme, i.e. if it had the form 
$\displaystyle{S(p)= \frac{i}{\left (p\!\!\!/-m \right)^{\sigma }}}$, 
with $\sigma \neq 1$, the identity (\ref{ward}) would \textit{not} be fulfilled.

\section{Fermion propagator in the Light-Front}

With the light-front coordinate transformations given in Appendix A, we can
find the corresponding fermionic propagator, beginning with the term
p\negthinspace \negthinspace \negthinspace /, as in (\ref{p barra fl}):

\[
p\!\!\!/=p_{\mu }\gamma ^{\mu }=\left( \gamma ^{+}p^{-}+\gamma
^{-}p^{+}\right) -\left( \overrightarrow{\gamma }_{\perp }\cdot 
\overrightarrow{p}_{\perp }\right) \text{,}
\]

then

\[
S(p)=\frac{i}{\left[ \left( \gamma ^{+}p^{-}+\gamma
^{-}p^{+}\right) -\left( \overrightarrow{\gamma }_{\perp }\cdot 
\overrightarrow{p}_{\perp }\right) -m\right] }\text{,}
\]

or, in another way, using $S(p)=\displaystyle{\frac{i\left( p\!\!\!/+m\right) }{p^{2}-m^{2}
}}$,

\begin{eqnarray}
&&\bigskip \left. S(p)=\frac{i\left[ \left( \gamma ^{+}p^{-}+\gamma
^{-}p^{+}\right) -\left( \overrightarrow{\gamma }_{\perp }\cdot 
\overrightarrow{p}_{\perp }\right) +m\right] }{p^{+}\left(
p^{-}-p_{on}\right) }\text{,}\right.   \nonumber \\
&&  \nonumber \\
&& \left. S(p)=\frac{i\left( p\!\!\!/_{on}+m\right) }{2p^{+}\left(
p^{-}-p_{on}\right) }+\frac{i\gamma ^{+}}{2p^{+}}\text{,}\right. 
\label{ward fl}
\end{eqnarray}

where

\[
p\!\!\!/_{on}=\left( \gamma ^{+}p_{on}+\gamma ^{-}p^{+}\right) -\left( 
\overrightarrow{\gamma }_{\perp }\cdot \overrightarrow{p}_{\perp }\right) 
\text{ ,}
\]

\[
p_{on}=\frac{p_{\perp }^{2}+m^{2}}{2p^{+}}\text{.}
\]

\section{The Ward Identity on the Light-Front}

There are two manners to test if the propagator (\ref{ward fl}) on
the light-front satisfy the Ward identity (\ref{ward}). The simplest and
most direct one is to do the derivatives $\displaystyle{\frac{\partial S_{_{{}}}^{-1}(p)}{
\partial p^{\mu }}}$ for each component and put them in (\ref{passo}):

\[
\frac{\partial S_{{}}^{-1}(p)}{\partial p^{+}}=-i\gamma ^{-}
\]

\[
\frac{\partial S_{{}}^{-1}(p)}{\partial p^{-}}=-i\gamma ^{+}
\]

\begin{equation}
\frac{\partial S_{{}}^{-1}(p)}{\partial p_{1,2}}=-i\gamma ^{1,2}\text{.}
\label{s-1 resultados}
\end{equation}

\[
\frac{\partial S(p)}{\partial p^{+}}=iS(p)\gamma ^{-}S(p)
\]

\[
\frac{\partial S(p)}{\partial p^{-}}=iS(p)\gamma ^{+}S(p)
\]

\begin{equation}
\frac{\partial S(p)}{\partial p_{1,2}}=iS(p)\gamma ^{1,2}S(p),
\label{s fl resultados}
\end{equation}
where $p_{1,2}=p_{\perp }$ and $\gamma ^{1,2}=\gamma ^{\perp }$ are the
transversal or perpendicular components.

Comparing (\ref{s fl resultados}) and (\ref{ward}), one verifies
that the Ward identity is satisfied if one includes the necessary factors
due to the change of coordinate system, or, in other words, considering the
Jacobian determinant of this transformation.

The second manner to test the identity on the light-front is working
explicitly with all the figures of (\ref{ward}). The details are presented
in Appendix B, and below we put the principal results:

\begin{equation}
\frac{\partial S(p)}{\partial p^{+}}=iS(p)\gamma ^{-}S(p)=\frac{%
-ip^{-}\left( p\!\!\!/+m\right) }{2\left[ p^{+}\left( p^{-}-p_{on}\right) %
\right] ^{2}}+\frac{i\gamma ^{-}}{2p^{+}\left( p^{-}-p_{on}\right) }\text{,}
\label{derivada +}
\end{equation}

\begin{equation}
\frac{\partial S(p)}{\partial p^{-}}=iS(p)\gamma ^{+}S(p)=\frac{-i\left(
p\!\!\!/_{on}+m\right) }{2p^{+}\left( p^{-}-p_{on}\right) ^{2}}\text{,}
\label{derivada -}
\end{equation}

\begin{equation}
\frac{\partial S(p)}{\partial p_{1,2}}=iS(p)\gamma ^{1,2}S(p)=\frac{%
ip_{1,2}\left( p\!\!\!/+m\right) }{2\left[ p^{+}\left( p^{-}-p_{on}\right) %
\right] ^{2}}+\frac{i\gamma ^{1,2}}{2p^{+}\left( p^{-}-p_{on}\right) }\text{,%
}  \label{derivada transversal}
\end{equation}

that is, again one corroborates the relations (\ref{s fl resultados}). An
important point here is that, using the simplified propagator $\displaystyle{S(p)=\frac{
i\left( p\!\!\!/_{on}+m\right) }{p^{+}\left( p^{-}-p_{on}\right) }}$ as some
authors do, the Ward Identity is \textit{not} fulfilled, as shown in
Appendix C.

\noindent

\section{Conclusions}

We have shown here that the Ward identity for the fermionic field in the light-front is preserved to guarantee that the charge renomalization constant depends solely on the photon renormalization constant, as it is expected. However, one important point emerges in our computation, and that is that the Ward identity in the light-front is valid provided the fermionic field propagator bears the relevant ``contact'' term piece, which is absent in the covariant propagator and its straightforward projection into light-front variables.

Our computation has demonstrated once again the significance of the light-front zero-mode contribution that the so-called ``contact'' term bears in it, without which Ward identity would be violated. Although the zero-mode term does not carry physical information, its non-vanishing contribution nonetheless is crucial to the validity of the Ward identity in the light-front formalism. In other words, ``contact'' term may not carry information from one space-time point to another in the light front, but contains relevant physical information needed to ensure the Ward identity, and therefore, for the correct charge renormalization.

\section{Appendix}

\subsection{Light-front Coordinates}

The Light-front is characterized by the null-plane $x^{+}=t+z=0$, which is
its time coordinate. All of the coordinates are set regarding this plane,
and one has new definitions of the scalar product, for example. The basic
relations on the light-front are

\begin{eqnarray}
&&\left. x^{+}=\frac{1}{\sqrt{2}}\left( x^{0}+x^{3}\right) \right.  
\nonumber \\
&&\left. x^{-}=\frac{1}{\sqrt{2}}\left( x^{0}-x^{3}\right) \right.  
\nonumber \\
&&\left. \overrightarrow{x}_{\perp }=x^{1}\overrightarrow{i}+x^{2}%
\overrightarrow{j}\right. \text{,}  \label{light front]}
\end{eqnarray}
so, the scalar product is given by

\begin{equation}
a^{\mu }b_{\mu }=\left( a^{+}b^{-}+a^{-}b^{+}\right) -\overrightarrow{a}%
_{\perp }\cdot \overrightarrow{b}_{\perp }\text{.}  \label{produto escalar]}
\end{equation}

Using (\ref{produto escalar]}), one can write the product $p\!\!\!/$ on the
light-front:

\begin{equation}
p\!\!\!/=p_{\mu }\gamma ^{\mu }=\left( \gamma ^{+}p^{-}+\gamma
^{-}p^{+}\right) -\left( \overrightarrow{\gamma }_{\perp }\cdot 
\overrightarrow{p}_{\perp }\right) \text{.}  \label{p barra fl}
\end{equation}

\subsection{Checking the Ward Identity}

In this Appendix, we show the details of the algebra necessary to arrive at (\ref{derivada +}-\ref
{derivada transversal}). In the first place, we list the numerous properties
that Dirac gama matrices in the light-front obey and should be used:

\begin{equation}
\begin{array}{cc}
\left. \gamma ^{+}\gamma ^{+}=\gamma ^{-}\gamma ^{-}=0\right.  & \left.
\gamma ^{1}\gamma ^{\pm }\gamma ^{2}+\gamma ^{2}\gamma ^{\pm }\gamma
^{1}=0\right.  \\ 
\left. \gamma ^{+}\gamma ^{-}\gamma ^{+}=2\gamma ^{+}\right.  & \left.
\gamma ^{\mp }\gamma ^{\pm }\gamma ^{1,2}+\gamma ^{1,2}\gamma ^{\pm }\gamma
^{\mp }=2\gamma ^{1,2}\right.  \\ 
\left. \gamma ^{-}\gamma ^{+}\gamma ^{-}=2\gamma ^{-}\right.  & \left.
\left\{ \left( \gamma _{\perp }p_{\perp }\right) ,\gamma ^{\pm }\right\}
=0\right.  \\ 
\left. \gamma ^{1}\gamma ^{\pm }\gamma ^{1}=\gamma ^{\pm }\right.  & \left.
\left\{ \gamma ^{+},\gamma ^{-}\right\} =2I\right.  \\ 
\left. \gamma ^{2}\gamma ^{\pm }\gamma ^{2}=\gamma ^{\pm }\right.  & \left.
\left( \gamma _{\perp }p_{\perp }\right) \gamma ^{\pm }\left( \gamma _{\perp
}p_{\perp }\right) =\left( p_{\perp }\right) ^{2}\gamma ^{\pm }\right.  \\ 
\left. \left\{ \gamma ^{\pm },\gamma ^{1,2}\right\} =0\right.  & \left.
\left\{ \left( \gamma _{\perp }p_{\perp }\right) ,\gamma ^{1,2}\right\}
=2p_{1,2}\right.  \\ 
\left. \gamma ^{\pm }\gamma ^{1,2}\gamma ^{\pm }=0\right.  & \left. \gamma
^{\pm }\gamma ^{\mp }\left( \gamma _{\perp }p_{\perp }\right) +\left( \gamma
_{\perp }p_{\perp }\right) \gamma ^{\mp }\gamma ^{\pm }=2\left( \gamma
_{\perp }p_{\perp }\right) \right.  \\ 
\left. \gamma ^{1}\gamma ^{1}=\gamma ^{2}\gamma ^{2}=-I\right.  & \left.
\gamma ^{\pm }\gamma ^{1,2}\left( \gamma _{\perp }p_{\perp }\right) +\left(
\gamma _{\perp }p_{\perp }\right) \gamma ^{1,2}\gamma ^{\pm }=2\gamma ^{\pm
}p_{1,2}\right.  \\ 
\left. \gamma ^{\pm }\gamma ^{1}\gamma ^{2}+\gamma ^{2}\gamma ^{1}\gamma
^{\pm }=0\right.  & \left. \gamma ^{+}\gamma ^{1,2}\gamma ^{-}+\gamma
^{-}\gamma ^{1,2}\gamma ^{+}=-2\gamma ^{1,2}\right.  \\ 
\left. \left\{ \gamma ^{1},\gamma ^{2}\right\} =0\right.  & \left. \left(
\gamma _{\perp }p_{\perp }\right) \gamma ^{1,2}\left( \gamma _{\perp
}p_{\perp }\right) =\mp \left( p_{1}\right) ^{2}\gamma ^{1,2}\pm \left(
p_{2}\right) ^{2}\gamma ^{1,2}-2p_{1}p_{2}\gamma ^{2,1}\right. 
\end{array}
\label{gamas}
\end{equation}

Next, some useful relations:

\begin{equation}
\frac{\partial p_{on}}{\partial p^{+}}=-\frac{p_{\perp }^{2}+m^{2}}{2\left(
p^{+}\right) ^{2}}=-\frac{p_{on}}{p^{+}}  \label{derivada pon}
\end{equation}

\begin{equation}
\frac{\partial p\!\!\!/_{on}}{\partial p^{+}}=-\gamma ^{+}\frac{p_{on}}{p^{+}%
}+\gamma ^{-}  \label{derivada p barra on}
\end{equation}

\begin{equation}
\frac{\partial p_{on}}{\partial p_{1}}=\frac{p_{1}}{p^{+}}.
\end{equation}

Remembering that the fermion propagator is $S(p)=\displaystyle\frac{i\left(
p\!\!\!/_{on}+m\right) }{p^{+}\left( p^{-}-p_{on}\right) }+\frac{i\gamma ^{+}
}{2p^{+}}$, the plus component derivative is

\[
\frac{\partial S(p)}{\partial p^{+}}=\frac{i\left( \frac{\partial
p\!\!\!/_{on}}{\partial p^{+}}\right) }{2p^{+}\left( p^{-}-p_{on}\right) }-%
\frac{i\left( p\!\!\!/_{on}+m\right) }{2\left( p^{+}\right) ^{2}\left(
p^{-}-p_{on}\right) }+\frac{i\left( p\!\!\!/_{on}+m\right) }{2p^{+}\left(
p^{-}-p_{on}\right) ^{2}}\left( \frac{\partial p_{on}}{\partial p^{+}}%
\right) -\frac{i\gamma ^{+}}{2\left( p^{+}\right) ^{2}}
\]

\[
\frac{\partial S(p)}{\partial p^{+}}=\frac{-i\gamma ^{+}\frac{p_{on}}{p^{+}}%
+i\gamma ^{-}}{2p^{+}\left( p^{-}-p_{on}\right) }-\frac{i\left(
p\!\!\!/_{on}+m\right) }{2\left( p^{+}\right) ^{2}\left( p^{-}-p_{on}\right) 
}-\frac{ip_{on}\left( p\!\!\!/_{on}+m\right) }{2\left( p^{+}\right)
^{2}\left( p^{-}-p_{on}\right) ^{2}}-\frac{i\gamma ^{+}}{2\left(
p^{+}\right) ^{2}}
\]

\[
\frac{\partial S(p)}{\partial p^{+}}=\frac{-i\gamma ^{+}\left( p^{-}\right)
^{2}-i\gamma ^{-}p^{+}p_{on}+i\left( \gamma _{\perp }p_{\perp }\right)
p^{-}-imp^{-}}{2\left[ p^{+}\left( p^{-}-p_{on}\right) \right] ^{2}}
\]

\[
\frac{\partial S(p)}{\partial p^{+}}=\frac{-ip^{-}\left( p\!\!\!/+m\right) +%
\frac{i}{2}p^{+}\gamma ^{-}\left( p^{-}-p_{on}\right) }{2\left[ p^{+}\left(
p^{-}-p_{on}\right) \right] ^{2}}
\]

\begin{equation}
\frac{\partial S(p)}{\partial p^{+}}=\frac{-ip^{-}\left( p\!\!\!/+m\right) }{%
2\left[ p^{+}\left( p^{-}-p_{on}\right) \right] ^{2}}+\frac{i\gamma ^{-}}{%
2p^{+}\left( p^{-}-p_{on}\right) }.  \label{resultado derivada positiva}
\end{equation}

\bigskip

Now, calculating the term $iS(p)\gamma ^{-}S(p)$ and exploiting the
properties of the gamma functions, we have

\[
=-i\left\{ \frac{\left( p\!\!\!/_{on}+m\right) \gamma ^{-}\left(
p\!\!\!/_{on}+m\right) }{4\left[ p^{+}\left( p^{-}-p_{on}\right) \right] ^{2}%
}+\frac{\left( p\!\!\!/_{on}+m\right) \gamma ^{-}\gamma ^{+}}{4\left(
p^{+}\right) ^{2}\left( p^{-}-p_{on}\right) }+\frac{\gamma ^{+}\gamma
^{-}\left( p\!\!\!/_{on}+m\right) }{4\left( p^{+}\right) ^{2}\left(
p^{-}-p_{on}\right) }+\frac{\gamma ^{+}\gamma ^{-}\gamma ^{+}}{4\left(
p^{+}\right) ^{2}}\right\} 
\]

\[
=-i\left\{ \frac{p\!\!\!/_{on}\gamma ^{-}p\!\!\!/_{on}+m\left\{
p\!\!\!/_{on},\gamma ^{-}\right\} +m^{2}\gamma ^{-}}{4\left[ p^{+}\left(
p^{-}-p_{on}\right) \right] ^{2}}+\frac{p\!\!\!/_{on}\gamma ^{-}\gamma
^{+}+\gamma ^{+}\gamma ^{-}p\!\!\!/_{on}+m\left\{ \gamma ^{+},\gamma
^{-}\right\} }{4\left( p^{+}\right) ^{2}\left( p^{-}-p_{on}\right) }+\frac{%
\gamma ^{+}}{2\left( p^{+}\right) ^{2}}\right\} 
\]

\[
=-i\left\{ \frac{2\gamma ^{+}\left( p_{on}\right) ^{2}-2p_{on}\left( \gamma
_{\perp }p_{\perp }\right) +\left( p_{\perp }\right) ^{2}\gamma
^{-}+2mp_{on}+m^{2}\gamma ^{-}}{4\left[ p^{+}\left( p^{-}-p_{on}\right) %
\right] ^{2}}+\frac{4\gamma ^{+}p_{on}-2\left( \gamma _{\perp }p_{\perp
}\right) +2m}{4\left( p^{+}\right) ^{2}\left( p^{-}-p_{on}\right) }+\frac{%
\gamma ^{+}}{2\left( p^{+}\right) ^{2}}\right\} 
\]

\[
=-i\left\{ \frac{2p^{-}p\!\!\!/-2\gamma ^{-}p^{+}\left( p^{-}-p_{on}\right)
+2mp^{-}}{4\left[ p^{+}\left( p^{-}-p_{on}\right) \right] ^{2}}\right\} 
\]

\begin{equation}
=\frac{-ip^{-}\left( p\!\!\!/+m\right) }{2\left[ p^{+}\left(
p^{-}-p_{on}\right) \right] ^{2}}+\frac{i\gamma ^{-}}{2p^{+}\left(
p^{-}-p_{on}\right) }\text{.}  \label{resultado produto positivo}
\end{equation}

One can the see that, from (\ref{resultado derivada positiva}) and (\ref
{resultado produto positivo}), $\displaystyle\frac{\partial S(p)}{\partial p^{+}}
=iS(p)\gamma ^{-}S(p)$.

For the minus component, the derivative is very simple,

\begin{equation}
\frac{\partial S(p)}{\partial p^{-}}=-\frac{i\left( p\!\!\!/_{on}+m\right) }{%
2p^{+}\left( p^{-}-p_{on}\right) ^{2}}  \label{resultado derivada negativa}
\end{equation}

And the term $iS(p)\gamma ^{+}S(p)$,

\[
=-i\left\{ \frac{\left( p\!\!\!/_{on}+m\right) \gamma ^{+}\left(
p\!\!\!/_{on}+m\right) }{4\left[ p^{+}\left( p^{-}-p_{on}\right) \right] ^{2}%
}\right\} 
\]

\[
=-i\left\{ \frac{2\gamma ^{-}\left( p^{+}\right) ^{2}-2p^{+}\left( \gamma
_{\perp }p_{\perp }\right) +2p^{+}p_{on}\gamma ^{+}+2mp^{+}}{4\left[
p^{+}\left( p^{-}-p_{on}\right) \right] ^{2}}\right\} 
\]

\begin{equation}
=\frac{-i\left( p\!\!\!/_{on}+m\right) }{2p^{+}\left( p^{-}-p_{on}\right)
^{2}}  \label{resultado produto negativo}
\end{equation}

and again one has $\displaystyle\frac{\partial S(p)}{\partial p^{-}}=iS(p)\gamma ^{+}S(p)$.

Finally, the derivative of the transversal components:

\[
\frac{\partial S(p)}{\partial p_{1}}=\frac{i\left( \frac{\partial
p\!\!\!/_{on}}{\partial p_{1}}\right) }{2p^{+}\left( p^{-}-p_{on}\right) }+%
\frac{i\left( p\!\!\!/_{on}+m\right) }{2p^{+}\left( p^{-}-p_{on}\right) ^{2}}%
\left( \frac{\partial p_{on}}{\partial p_{1}}\right) 
\]

\[
\frac{\partial S(p)}{\partial p_{1}}=\frac{i\left( \gamma ^{+}\frac{p_{1}}{%
p^{+}}+\gamma ^{1}\right) }{2p^{+}\left( p^{-}-p_{on}\right) }+\frac{%
ip_{1}\left( p\!\!\!/_{on}+m\right) }{\left( p^{+}\right) ^{2}\left(
p^{-}-p_{on}\right) ^{2}}
\]

\[
\frac{\partial S(p)}{\partial p_{1}}=\frac{ip_{1}\left[ \gamma
^{+}p^{-}+\gamma ^{-}p^{+}-\left( \gamma _{\perp }p_{\perp }\right) +m\right]
+i\gamma ^{1}p^{+}\left( p^{-}-p_{on}\right) }{2\left[ p^{+}\left(
p^{-}-p_{on}\right) \right] ^{2}}
\]

\begin{equation}
\frac{\partial S(p)}{\partial p_{1}}=\frac{ip_{1}\left( p\!\!\!/+m\right) }{2%
\left[ p^{+}\left( p^{-}-p_{on}\right) \right] ^{2}}+\frac{i\gamma ^{1}}{%
2p^{+}\left( p^{-}-p_{on}\right) }\text{.}
\label{resultado derivada transversal}
\end{equation}

The term $iS(p)\gamma ^{1}S(p)$ is very laborious and almost all of the gamma
matrices properties must be used:

\[
=-i\left\{ \frac{p\!\!\!/_{on}\gamma ^{1}p\!\!\!/_{on}+m\left\{
p\!\!\!/_{on},\gamma ^{1}\right\} +m^{2}\gamma ^{1}}{4\left[ p^{+}\left(
p^{-}-p_{on}\right) \right] ^{2}}+\frac{p\!\!\!/_{on}\gamma ^{1}\gamma
^{+}+\gamma ^{+}\gamma ^{1}p\!\!\!/_{on}+m\left\{ \gamma ^{1},\gamma
^{-}\right\} }{4\left( p^{+}\right) ^{2}\left( p^{-}-p_{on}\right) }+\frac{%
\gamma ^{+}\gamma ^{1}\gamma ^{+}}{4\left( p^{+}\right) ^{2}}\right\} 
\]

\begin{eqnarray*}
&&\left. =-i\left\{ \frac{-2\gamma ^{1}p^{+}p_{on}-2\gamma
^{+}p_{1}p_{on}-2\gamma ^{-}p^{+}p_{1}-\gamma ^{1}\left( p_{1}\right)
^{2}+\gamma ^{1}\left( p_{2}\right) ^{2}-2\gamma ^{2}p_{1}p_{2}+m^{2}\gamma
^{1}-2mp_{1}}{4\left[ p^{+}\left( p^{-}-p_{on}\right) \right] ^{2}}+\right.
\right.  \\
&&\left. \left. +\frac{2\gamma ^{+}p_{1}-2\gamma ^{1}p^{+}}{4\left(
p^{+}\right) ^{2}\left( p^{-}-p_{on}\right) }\right\} \right. 
\end{eqnarray*}

\[
=-i\left\{ \frac{-2p_{1}\left[ \gamma ^{+}p^{-}+\gamma ^{-}p^{+}-\left(
\gamma _{\perp }p_{\perp }\right) +m\right] +\gamma ^{1}\left( p_{1}\right)
^{2}+\gamma ^{1}\left( p_{2}\right) ^{2}-2\gamma ^{1}p^{+}p^{-}+m^{2}\gamma
^{1}}{4\left[ p^{+}\left( p^{-}-p_{on}\right) \right] ^{2}}\right\} 
\]

\[
=-i\left[ \frac{-2p_{1}\left( p\!\!\!/+m\right) +\gamma ^{1}\left(
p_{1}^{2}+p_{2}^{2}+m^{2}-2p^{+}p^{-}\right) }{4\left[ p^{+}\left(
p^{-}-p_{on}\right) \right] ^{2}}\right] 
\]

\[
=-i\left[ \frac{-2p_{1}\left( p\!\!\!/+m\right) -2\gamma ^{1}\left(
p^{+}p^{-}-p^{+}p_{on}\right) }{4\left[ p^{+}\left( p^{-}-p_{on}\right) %
\right] ^{2}}\right] 
\]

\begin{equation}
=\frac{ip_{1}\left( p\!\!\!/+m\right) }{2\left[ p^{+}\left(
p^{-}-p_{on}\right) \right] ^{2}}+\frac{i\gamma ^{1}}{2p^{+}\left(
p^{-}-p_{on}\right) }  \label{resultado produto transversal}
\end{equation}

and from (\ref{resultado derivada transversal}) and (\ref{resultado produto
transversal}), one has $\displaystyle\frac{\partial S(p)}{\partial p_{1,2}}=iS(p)\gamma
^{1,2}S(p)$.

\subsection{The Ward Identity for the propagator without contact term}

Here we work on the Ward Identity for the simplified propagator $S(p)=\displaystyle\frac{
i\left( p\!\!\!/_{on}+m\right) }{2p^{+}\left( p^{-}-p_{on}\right) }$.

For the minus component, the derivative is the same as the one obtained before,

\begin{equation}
\frac{\partial S(p)}{\partial p^{-}}=\frac{-i\left( p\!\!\!/_{on}+m\right) }{
2p^{+}\left( p^{-}-p_{on}\right) ^{2}}\text{;}
\label{derivada negativa modificada}
\end{equation}

and the term $iS(p)\gamma ^{+}S(p)$ is equal too, because in the other case,
the terms $\displaystyle\frac{i\gamma ^{+}}{2p^{+}}$ do not contribute due to the
property $\gamma ^{+}\gamma ^{+}=0$:

\begin{equation}
iS(p)\gamma ^{+}S(p)=\frac{-i\left( p\!\!\!/_{on}+m\right) }{2p^{+}\left(
p^{-}-p_{on}\right) ^{2}}\text{,}  \label{produto negativo modificado}
\end{equation}

so, for the negative component, the Ward identity \textit{is} satisfied

\[
\frac{\partial S(p)}{\partial p^{-}}=iS(p)\gamma ^{+}S(p)\text{.}
\]

For the plus component, one has

\[
\frac{\partial S(p)}{\partial p^{+}}=\frac{-i\gamma ^{+}\frac{p_{on}}{p^{+}}%
+i\gamma ^{-}}{2p^{+}\left( p^{-}-p_{on}\right) }-\frac{i\left(
p\!\!\!/_{on}+m\right) }{2\left( p^{+}\right) ^{2}\left( p^{-}-p_{on}\right) 
}-\frac{ip_{on}\left( p\!\!\!/_{on}+m\right) }{2\left( p^{+}\right)
^{2}\left( p^{-}-p_{on}\right) ^{2}}
\]

\[
\frac{\partial S(p)}{\partial p^{+}}=\frac{ip^{+}p^{-}\gamma
^{-}-p^{+}p_{on}\gamma ^{-}-p^{-}p_{on}\gamma ^{+}+\left( p_{on}\right)
^{2}\gamma ^{+}-p^{-}\left( p\!\!\!/_{on}+m\right) }{2\left[ p^{+}\left(
p^{-}-p_{on}\right) \right] ^{2}}
\]

\begin{equation}
\frac{\partial S(p)}{\partial p^{+}}=\frac{-ip^{-}\left(
p\!\!\!/_{on}+m\right) }{2\left[ p^{+}\left( p^{-}-p_{on}\right) \right] ^{2}%
}+\frac{i\gamma ^{-}}{2p^{+}\left( p^{-}-p_{on}\right) }-\frac{i\gamma
^{+}p_{on}}{2\left( p^{+}\right) ^{2}\left( p^{-}-p_{on}\right) }\text{;}
\label{derivada positiva modificada}
\end{equation}

And the term $iS(p)\gamma ^{-}S(p)$,

\[
i\left[ \frac{i\left( p\!\!\!/_{on}+m\right) }{2p^{+}\left(
p^{-}-p_{on}\right) }\right] \gamma ^{-}\left[ \frac{i\left(
p\!\!\!/_{on}+m\right) }{2p^{+}\left( p^{-}-p_{on}\right) }\right] =-i\frac{%
\left( p\!\!\!/_{on}+m\right) \gamma ^{-}\left( p\!\!\!/_{on}+m\right) }{4%
\left[ p^{+}\left( p^{-}-p_{on}\right) \right] ^{2}}
\]

\[
\left. =-i\left\{ \frac{2\gamma ^{+}\left( p_{on}\right) ^{2}\gamma
^{+}-2p_{on}\left( \gamma _{\perp }p_{\perp }\right) +\left( p_{\perp
}\right) ^{2}\gamma ^{-}+2mp_{on}+m^{2}\gamma ^{-}}{4\left[ p^{+}\left(
p^{-}-p_{on}\right) \right] ^{2}}\right\} \right. 
\]

\[
=-i\left\{ \frac{2p^{-}\left( p\!\!\!/_{on}+m\right) +\left[ -2\gamma
^{+}p_{on}-2\gamma ^{-}p^{+}+2\left( \gamma _{\perp }p_{\perp }\right) -2m%
\right] \left( p^{-}-p_{on}\right) }{4\left[ p^{+}\left( p^{-}-p_{on}\right) %
\right] ^{2}}\right\} 
\]

\begin{equation}
=\frac{-ip^{-}\left( p\!\!\!/_{on}+m\right) }{2\left[ p^{+}\left(
p^{-}-p_{on}\right) \right] ^{2}}+\frac{i\gamma ^{-}}{2p^{+}\left(
p^{-}-p_{on}\right) }+\frac{i\gamma ^{+}p_{on}}{2\left( p^{+}\right)
^{2}\left( p^{-}-p_{on}\right) }-\frac{\left[ \left( \gamma _{\perp
}p_{\perp }\right) -m\right] }{2\left( p^{+}\right) ^{2}\left(
p^{-}-p_{on}\right) }\text{;}  \label{produto positivo modificado}
\end{equation}

and because of the presence of the last term and the wrong signal of the
third, one has

\[
\frac{\partial S(p)}{\partial p^{+}}\neq iS(p)\gamma ^{-}S(p)\text{.}
\]

For the transversal components, the derivative is the same as obtained before,

\begin{equation}
\frac{\partial S(p)}{\partial p_{1}}=\frac{ip_{1}\left( p\!\!\!/+m\right) }{2%
\left[ p^{+}\left( p^{-}-p_{on}\right) \right] ^{2}}-\frac{i\gamma ^{1}}{%
2p^{+}\left( p^{-}-p_{on}\right) }\text{.}
\label{derivada transversal  modificada}
\end{equation}

And the term $iS(p)\gamma ^{1}S(p)$,

\[
i\left[ \frac{i\left( p\!\!\!/_{on}+m\right) }{2p^{+}\left(
p^{-}-p_{on}\right) }\right] \gamma ^{1}\left[ \frac{i\left(
p\!\!\!/_{on}+m\right) }{2p^{+}\left( p^{-}-p_{on}\right) }\right] =-i\frac{%
\left( p\!\!\!/_{on}+m\right) \gamma ^{-}\left( p\!\!\!/_{on}+m\right) }{4%
\left[ p^{+}\left( p^{-}-p_{on}\right) \right] ^{2}}
\]

\begin{eqnarray*}
&&\left. =-i\left\{ \frac{-2p_{1}\left[ \gamma ^{+}p^{-}+\gamma
^{-}p^{+}-\left( \gamma _{\perp }p_{\perp }\right) +m\right] +2\gamma
^{+}p^{-}p_{1}+\gamma ^{1}\left( p_{1}\right) ^{2}+}{4\left[ p^{+}\left(
p^{-}-p_{on}\right) \right] ^{2}}\right. \right.  \\
&&\left. \left. \frac{-2\gamma ^{1}p^{+}p_{on}-2\gamma
^{+}p_{1}p_{on}+\gamma ^{1}\left( p_{2}\right) ^{2}+m^{2}\gamma ^{1}}{4\left[
p^{+}\left( p^{-}-p_{on}\right) \right] ^{2}}\right\} \right. 
\end{eqnarray*}

\begin{equation}
=\frac{ip_{1}\left( p\!\!\!/+m\right) }{2\left[ p^{+}\left(
p^{-}-p_{on}\right) \right] ^{2}}-\frac{\gamma ^{+}p_{1}}{2\left(
p^{+}\right) ^{2}\left( p^{-}-p_{on}\right) }\text{,}
\label{produto transversal  modificado}
\end{equation}

then, comparing (\ref{derivada transversal modificada}) and (\ref{produto
transversal modificado}), one has $\displaystyle\frac{\partial S(p)}{\partial p_{1,2}}
\neq iS(p)\gamma ^{1,2}S(p)$.

\bigskip

\bigskip

\textbf{Acknowledgments:} J.D. Bolzan thanks CNPq for financial support.
J.H.O. Sales thanks FAPESP and the hospitality of the Institute for
Theoretical Physics, UNESP, where part of this work has been performed.

\end{document}